\documentclass[reprint]{JASA-EL}

\usepackage{hyperref} 
\usepackage{fix-cm}
\begin{document}

%% Version in square brackets will appear at top right of all pages except title page:
\title[JASA-EL/Sample JASA-EL Article]{Deep-Learning-Assisted Full Waveform Inversion for Attenuation Reconstruction in Ultrasound Computed Tomography}
\author{Mingrui Liu}
\email{ml132@illinois.edu}
\author{Michael L. Oelze}
\altaffiliation{Also at: Carle Illinois College of Medicine, University of Illinois at Urbana-Champaign, Urbana, IL 61801 USA}
\correspondingauthor
\email{oelze@illinois.edu}
\affiliation{Department of Electrical and Computer Engineering, University of Illinois at Urbana-Champaign, Urbana, IL 61801 USA}
\affiliation{Beckman Institute for Advanced Science and Technology, University of Illinois at Urbana-Champaign, Urbana, IL 61801 USA}

\date{\today} 
\preprint{Author, JASA-EL}  %if you want want this message to appear in upper right corner of title page

\begin{abstract}
Acoustic attenuation is more challenging to reconstruct and has received less attention than sound speed in ultrasound computed tomography using full-waveform inversion (FWI). In this study, a deep-learning (DL)-assisted FWI framework was proposed for attenuation reconstruction based on our previous work. Some numerical experiments were conducted to investigate the strategy and the effectiveness of using DL to help accelerate and improve the reconstruction quality of attenuation imaging in FWI. The results indicated that DL saved computational cost by 28\% and improved attenuation reconstruction accuracy by 70\% when combined with sufficiently accurate sound speed estimates. The result supports the potential of DL initialization to improve the computational efficiency and reconstruction quality of joint SOS and attenuation FWI.
\end{abstract}

%% pacs numbers not used

\maketitle

%  End of title page for Preprint option --------------------------------- %

\section{Introduction}
Ultrasound computed tomography (USCT) can provide high quality and quantitative images of the breast without speckle artifacts \cite{Wang2015Waveform}. It has great potential for 3D quantitative imaging of acoustic breast tissue properties because it can record reflection, acoustic attenuation, and speed of sound (SOS) images simultaneously \cite{wiskin20173, wiskin2020full}. In the past two decades, the full wave inversion (FWI) \cite{malikqt2018,Liva2017Time} approach has been adopted in USCT, providing much higher resolution compared to bent-ray algorithms \cite{li2009vivo}.

Most USCT related studies have focused on improving SOS reconstruction \cite{jeong2024investigating,wiskin2022imaging,wiskin2023whole}, but few studies have investigated attenuation imaging. The main reason is that attenuation reconstruction is a more difficult task compared to SOS using the FWI method \cite{liu2025angular,sarno2025}. Attenuation is a numerically poorly conditioned term in the optimization problem because it is encoded in the imaginary part of the wavenumber \cite{jiang2018identifying}. Also, it is more sensitive to noise in real scenarios, as attenuation leads to the change in the signal amplitude that is not only dependent on attenuation, but also refraction, focusing (diffraction), and scattering.

In this study, we investigated the effectiveness of deep-learning (DL) based attenuation initialization for FWI in USCT. Building upon our previous work on DL-based attenuation imaging \cite{liudlatten}, we evaluated whether DL initialization could improve attenuation reconstruction accuracy and computational efficiency compared with conventional uniform initialization. In particular, we examined how the accuracy of the initial SOS estimate influenced the benefits of DL-assisted inversion. Through numerical experiments with different initialization strategies and levels of SOS preconditioning, we demonstrated the conditions under which DL initialization provided meaningful improvements in attenuation reconstruction.

\section{Methods}
\subsection{Numerical experiment setup}
\begin{figure}[!ht]
\centerline{\includegraphics[width=\columnwidth]{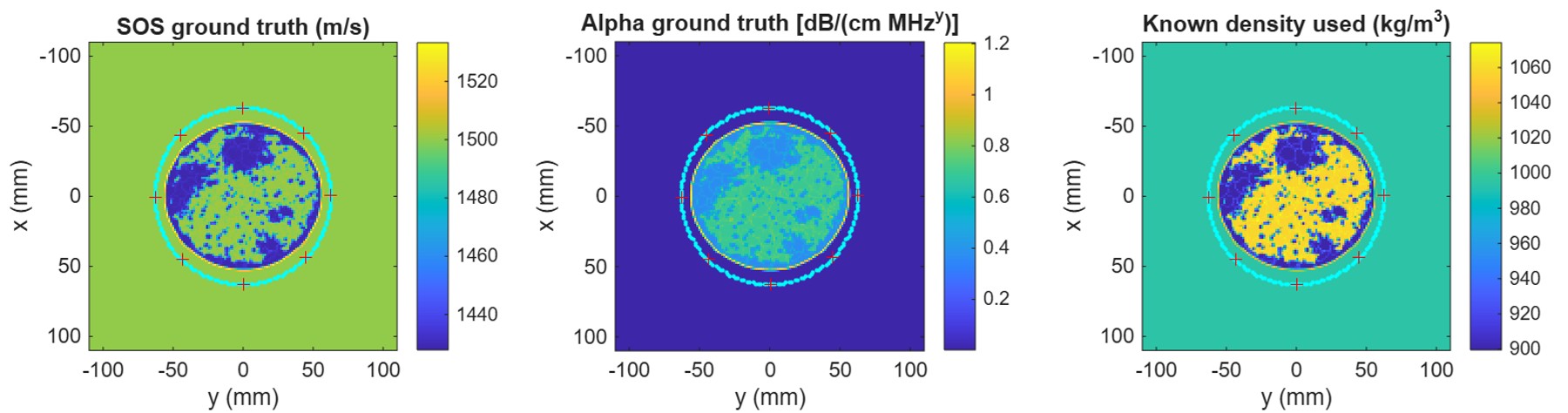}}
\caption{FWI simulation experiment settings.}
\label{fig1}
\end{figure}

The dataset used and a description of the generation of the datset are decribed in a previous work \cite{li20213}\cite{liudlatten}. For this numerical experiment, one slice from breast phantom C was selected from the dataset \cite{liudlatten}. The ground truth SOS, attenuation, density map, and the experiment configuration can be found in Fig. \ref{fig1}. Eight independent transmit locations and 192 receive locations were placed on a circular ring. Transmit and receive sites were individual grid points, without finite aperture, directional response, or element averaging.

The unknown reconstruction ROI contained 5402 active pixels inside the breast region. Outside the breast phantom ROI, SOS and attenuation were set to their water values. The active scaled parameter vector for a joint inversion was:
\begin{equation} \mathbf{q}=\begin{bmatrix}\mathbf{q}_c\\\mathbf{q}_a\end{bmatrix},\quad
\mathbf{q}_c=P_\Omega[(c-1500)/100],\quad
\mathbf{q}_a=P_\Omega a. \label{app1}\end{equation}
where \(P_\Omega\) was the operator that extracted and vectorized the ROI pixels. Thus \(\mathbf{q}\in\mathbb{R}^{10804}\).

\subsection{Deep Learning Initialization for Joint Sound-Speed and Attenuation Inversion}
The DL attenuation result can be used as the initialization for Joint SOS and attenuation reconstruction in FWI. Our previous work \cite{liudlatten} generated the predicted result by training from 60-angles backscattered signal and the SOS map. 

\subsection{Solver Definition}
In this section, we briefly describe the FWI solver. 
\subsubsection{FFT operators and absorption
coefficients}

Let \(\mathcal F\) denote the two dimensional discrete Fourier transform. Frequency vectors were
\(k_x,k_y=2\pi[0,\ldots,N/2-1,-N/2,\ldots,-1]/L\), with
\(K=(k_x^2+k_y^2)^{1/2}\). The radial wavenumber \(K\) was computed from the original frequency grids, while the Nyquist component of each directional wavenumber was set to zero only when constructing the corresponding spatial derivative operator. Define:
 \begin{equation} 
\kappa(K)=\frac{\sin(c_{\mathrm{ref}}K\Delta t/2)}{c_{\mathrm{ref}}K\Delta t/2},\quad
 c_{\mathrm{ref}}=1500\ \mathrm{m/s},\quad \kappa(0)=1.\label{app2}\end{equation}

The derivative operators were:
 \begin{equation} D_jv=\operatorname{Re}\{\mathcal F^{-1}[\mathrm{i}k_j\kappa(K)\mathcal F(v)]\},\quad j=x,y.\label{app3}\end{equation}

Under the discrete Euclidean inner product, \(D_j^T=-D_j\), which was used in discrete adjoint gradient calculation.

The fractional operators were defined as:
\begin{equation}
\mathcal Tv=\operatorname{Re}\{\mathcal F^{-1}[m_T(K)\mathcal F(v)]\},\quad
 m_T(K)=\begin{cases}K^{y-2},&K>0,\\0,&K=0,\end{cases}\label{app4}\end{equation}
\begin{equation}
\mathcal Ev=\operatorname{Re}\{\mathcal F^{-1}[K^{y-1}\mathcal F(v)]\}.\label{app5}\end{equation}

The coefficient conversion and local absorption and dispersion coefficients were as follows, where \(y=1.5\) in this implementation:
\begin{equation}
B_y=\frac{100}{20\log_{10}(e)(2\pi10^6)^y},\quad
 \tau=-2B_yac^{y-1},\quad \eta=2B_yac^y\tan(\pi y/2).\label{app6}\end{equation}

\subsubsection{Forward update}

All products, divisions, and powers below are element-wise. Let \(\widetilde\rho=\rho_0/(1000\ \mathrm{kg/m^3})\) be the known normalized density. The solver states were the computational velocity components \(u,v\), compression related variable \(r\), and pressure \(p\). All states were initialized to zero for each shot. The fixed array $d(\mathbf{x})$ denotes boundary damping.

For frequency hop $b$, the source pulse with center frequency $f_b$ was sampled as follows:
\begin{equation}
\begin{aligned}
s_{b,n}
&=\frac{1}{c_w^2}
\sin\left[2\pi f_b(t_n-t_{0,b})\right]
\exp\left[-\left(\frac{t_n-t_{0,b}}{w_b}\right)^2\right],\\
t_n&=(n-1)\Delta t,\qquad
t_{0,b}=\frac{2.5}{f_b},\qquad
w_b=\frac{0.7}{f_b},
\end{aligned}
\label{app_source}
\end{equation}
where $c_w=1500$~m/s and $f_b\in\{0.08,0.15,0.25\}$~MHz.
This fixed numerical source normalization was identical in data generation and inversion. With $e_s$ denoting the unit vector at transmit location $s$, and \(\mathcal R\) samples pressure at the receivers, one time step was computed as follows:

\begin{equation}
u^n=d\odot\left(u^{n-1}-\Delta t\,\widetilde\rho^{-1}\odot D_xp^{n-1}\right)\label{app7}\end{equation}

\begin{equation}v^n=d\odot\left(v^{n-1}-\Delta t\,\widetilde\rho^{-1}\odot D_yp^{n-1}\right),\label{app8}\end{equation}

\begin{equation}h^n=\widetilde\rho\odot(D_xu^n+D_yv^n),\quad
 r^n=d\odot(r^{n-1}-\Delta t\,h^n)+e_s s_{b,n}, \label{app9}\end{equation}

\begin{equation}p^n=c^2\odot\left[r^n+\tau\odot\mathcal Th^n-\eta\odot\mathcal Er^n\right],\quad
 F_{b,s}(c,a)_{:,n}=\mathcal Rp^n. \label{app10}\end{equation}

\subsubsection{Loss function}
For every band and shot, observations were generated as
\(d_{b,s}=F_{b,s}(c_{\mathrm{true}},a_{\mathrm{true}})\) with the true known density. The same operators, numerical precision, source, and acquisition were used during inversion.

Define the band energy \(Z_b=\max(\sum_s\|d_{b,s}\|_F^2,\epsilon_{\mathrm{mach}})\). The objective at band \(b\) was:
\begin{equation}
J_b(\mathbf q)=\frac{1}{2Z_b}\sum_{s=1}^{8}
 \|F_{b,s}(c(\mathbf q),a(\mathbf q))-d_{b,s}\|_F^2
 .\label{app11}\end{equation}

The sums include every receiver and time sample. \(Z_b\) is one normalization for the entire band.

\subsubsection{Discrete adjoint gradient}
Let \(\mathbf e_{b,s,n}=[F_{b,s}(c,a)_{:,n}-d_{b,s,:,n}]/Z_b\). The Euclidean discrete data gradient is the sum of Jacobian transpose actions,
\begin{equation}
\nabla_{(c,a)}J_b=\sum_s[\partial_{(c,a)}F_{b,s}]^T\mathbf e_{b,s}.\label{app12}\end{equation}

During gradient evaluation, \(h^n\) and \(r^n\) were stored for all time samples. Define \(t^n=\mathcal Th^n\), \(z^n=\mathcal Er^n\) and \(\psi^n=r^n+\tau\odot t^n-\eta\odot z^n\). At each \(n=N_t,\ldots,1\), the contribution from the receiver residual was added to the pressure adjoint:
\begin{equation}
\widehat{\bar p}=\bar p+\mathcal R^T\mathbf e_{b,s,n}.\label{app13}\end{equation}

Bars denote adjoint variables, which accumulate derivatives of the objective with respect to the intermediate wavefield states. All adjoint variables were initialized to zero.

For each shot, the parameter-gradient arrays were initialized to zero. At each reverse time step, the update was as follows:
\begin{equation} g_c^{(s)}\mathrel{+}=\widehat{\bar p}\odot\left[
 2c\odot\psi^n+c^2\odot\left((y-1)\frac{\tau}{c}\odot t^n-y\frac{\eta}{c}\odot z^n\right)\right], \label{app:gc}\end{equation}

\begin{equation} g_a^{(s)}\mathrel{+}=\widehat{\bar p}\odot c^2\odot
 \left[-2B_yc^{y-1}\odot t^n-2B_yc^y\tan(\pi y/2)\odot z^n\right]. \label{app:ga}\end{equation}

\begin{equation} z_p=c^2\odot\widehat{\bar p},\quad
 b_r=\bar r+z_p-\mathcal E(\eta\odot z_p),\label{app_rev1}
\end{equation}

\begin{equation} b_h=\mathcal T(\tau\odot z_p)-\Delta t\,d\odot b_r,\quad
 \bar r_{\mathrm{prev}}=d\odot b_r, \label{app_rev2}
\end{equation}

\begin{equation} b_u=\bar u-D_x(\widetilde\rho\odot b_h),\quad
 b_v=\bar v-D_y(\widetilde\rho\odot b_h), \label{app_rev3}
\end{equation}

\begin{equation} \bar p_{\mathrm{prev}}=\Delta t\left[
 D_x\left(\frac{d\odot b_u}{\widetilde\rho}\right)
 +D_y\left(\frac{d\odot b_v}{\widetilde\rho}\right)\right], \label{app_rev4}
\end{equation}

\begin{equation} \bar u_{\mathrm{prev}}=d\odot b_u,\quad
 \bar v_{\mathrm{prev}}=d\odot b_v.\label{app_rev5}
\end{equation}

At the end of each reverse time step, the sensitivities
$(\bar p,\bar r,\bar u,\bar v)$ were replaced by $(\bar p_{\mathrm{prev}},\bar r_{\mathrm{prev}}, \bar u_{\mathrm{prev}},\bar v_{\mathrm{prev}})$ before proceeding to the preceding time step.

After accumulation over all time steps and shots, the gradient with respect to the active scaled parameters was
\begin{equation}
\nabla_{\mathbf q}J_b=
\begin{bmatrix}
100P_\Omega\displaystyle\sum_s g_c^{(s)}\\[3pt]
P_\Omega\displaystyle\sum_s g_a^{(s)}
\end{bmatrix}.
\label{app_scaled_gradient}
\end{equation}
 
\subsubsection{Optimizer}
The scaled parameters were updated using a custom projected limited-memory BFGS (L-BFGS) optimizer \cite{liu1989limited}. Sound speed and attenuation were constrained to $[1300,1800]$~m/s and $[0,3]$ $\mathrm{dB}/(\mathrm{cm}\cdot\mathrm{MHz}^{1.5})$, respectively. Frequency hopping was performed using center frequencies of 0.08, 0.15 and 0.25 MHz, in ascending order. The parameter maps obtained at each frequency were used to initialize the next frequency, while the L-BFGS history was reset.

\subsection{Experiment A: Initialization Comparison}
First, five different experiment combinations were conducted, corresponding to different attenuation and SOS initializations, which can be found in Table \ref{tab:inversion_combinations1}. For all the experimental settings, the uniform initialization for SOS was 1520 m/s, and 0.4 dB/(cm·MHz\(^{1.5}\)) for attenuation. From Table \ref{tab:inversion_combinations1}, the sequentialUniform (Combination E) refers to the case where the SOS initialization came from the reconstructed SOS generated from JointUniform (Combination C).

\begin{table}[!htbp]
\centering
\caption{Initialization strategies for attenuation reconstruction experiment A.}
\label{tab:inversion_combinations1}
\small
\renewcommand{\arraystretch}{1.2}

\begin{tabular*}{\textwidth}
{@{\extracolsep{\fill}}lll@{}}
\hline\hline
\textbf{Combination} &
\textbf{Attenuation initialization} &
\textbf{SOS initialization} \\
\hline
A: OracleAlpha       & Uniform       & Ground truth \\
B: OracleSOS         & Ground truth  & Uniform \\
C: JointUniform      & Uniform       & Uniform \\
D: JointDL           & DL prediction & Uniform \\
E: SequentialUniform & Uniform       & From C \\
\hline\hline
\end{tabular*}
\end{table}

\subsection{Experiment B: DL Initialization with Improved SOS}
The second experiment started with an improved SOS initialization, which used the SOS map from JointUniform from Table. \ref{tab:inversion_combinations1} in Experiment A. Then, two different combinations were evaluated. One was uniform attenuation initialization, which was called JointUniformWarm, and the other one was the DL attenuation initialization, which was called JointDLWarm. In both combinations, SOS was jointly updated with JointUniform initialization.

\subsection{Experiment C: SOS Preconditioning Study}
Finally, To investigate whether DL-based attenuation initialization could reduce the SOS preconditioning required for subsequent joint inversion, a paired experiment was performed. In the preconditioning stage, sound speed was initialized uniformly to 1520 m/s, while the attenuation coefficient was fixed at a uniform value of 0.4 dB/(cm·MHz\(^{1.5}\)). Sound speed alone was updated using frequency continuation with pulse center frequencies of 0.08, 0.15, and 0.25 MHz, allowing up to 50 accepted iterations per frequency band. Intermediate SOS maps were retained after 0, 30, 75, 100, 110, 120, 130, 140, and 150 cumulative accepted iterations. Each retained sound-speed map was subsequently used to initialize two joint inversions. In the first, attenuation was initialized uniformly to 0.4 dB/(cm·MHz\(^{1.5}\)); in the second, it was initialized using the DL prediction. Both sound speed and attenuation were updated simultaneously in both cases. 

\section{Results}
\subsection{Experiment A: Initialization Comparison}
\begin{figure}[!ht]
\centerline{\includegraphics[width=\columnwidth]{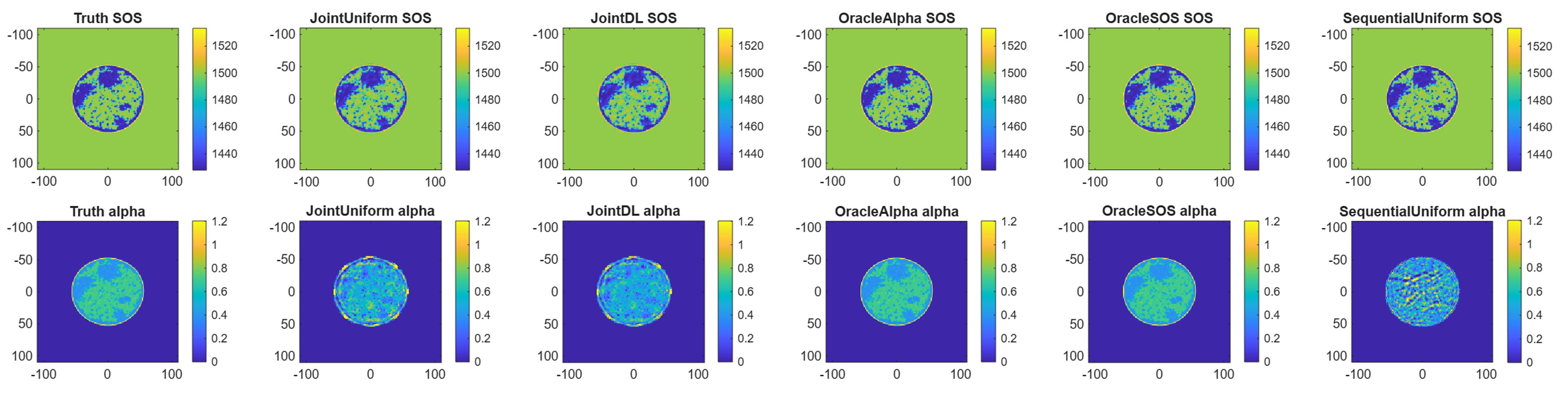}}
\caption{Reconstruction results with different initialization strategies for FWI for experiment A.}
\label{fig:exp1_recon}
\end{figure}

The results of five different combinations can be found in Fig. \ref{fig:exp1_recon} and Table \ref{tab:initial_inversion_results}. There are three findings from this first experiment. First, from the results of the combination OracleAlpha, with ground truth SOS, the attenuation can be reconstructed almost perfectly, proving the correctness of this simulation forward operator and adjoint gradient calculations. Second, it is more difficult to reconstruct attenuation compared to SOS. From the result of JointUniform and JointDL, the SOS image is improved compared to attenuation, which shows much clearer structural information. Third, a good initialization for attenuation does not result in improved performance if SOS has a large bias. Comparing the results of JointDL with the results of JointUniform, for both cases, the RMSE of attenuation did not converge and had a large RMSE in the end, indicating that the DL attenuation initialization did not result in improved SOS reconstruction.

\begin{table}[!htbp]
\centering
\caption{Reconstruction accuracy, relative RF residual, and computational cost for different inversion strategies.}
\label{tab:initial_inversion_results}
\small
\renewcommand{\arraystretch}{1.2}
\setlength{\tabcolsep}{4pt}

\begin{tabular*}{\textwidth}
{@{\extracolsep{\fill}}lccccc@{}}
\hline\hline
\textbf{Combination} &
\textbf{SOS RMSE} &
\textbf{Alpha RMSE} &
\textbf{RF residual} &
\textbf{Forward} &
\textbf{Adjoint} \\
\hline
JointUniform
    & 4.00 & 0.249 & 0.011 & 1464 & 744 \\
JointDL
    & 5.014 & 0.245 & 0.015 & 1464 & 744 \\
OracleAlpha
    & 0 &  $6.00\times10^{-4}$ & $6.27\times10^{-6}$ & 1488 & 744 \\
OracleSOS
    & 1.165 & 0 & 0.003 & 1472 & 744 \\
SequentialUniform
    & 1.380 & 0.270 & 0.007 & 2928 & 1488 \\
\hline\hline
\end{tabular*}
\end{table}

\subsection{Experiment B: DL Initialization with Improved SOS}

\begin{figure}[!ht]
\centerline{\includegraphics[width=\columnwidth]{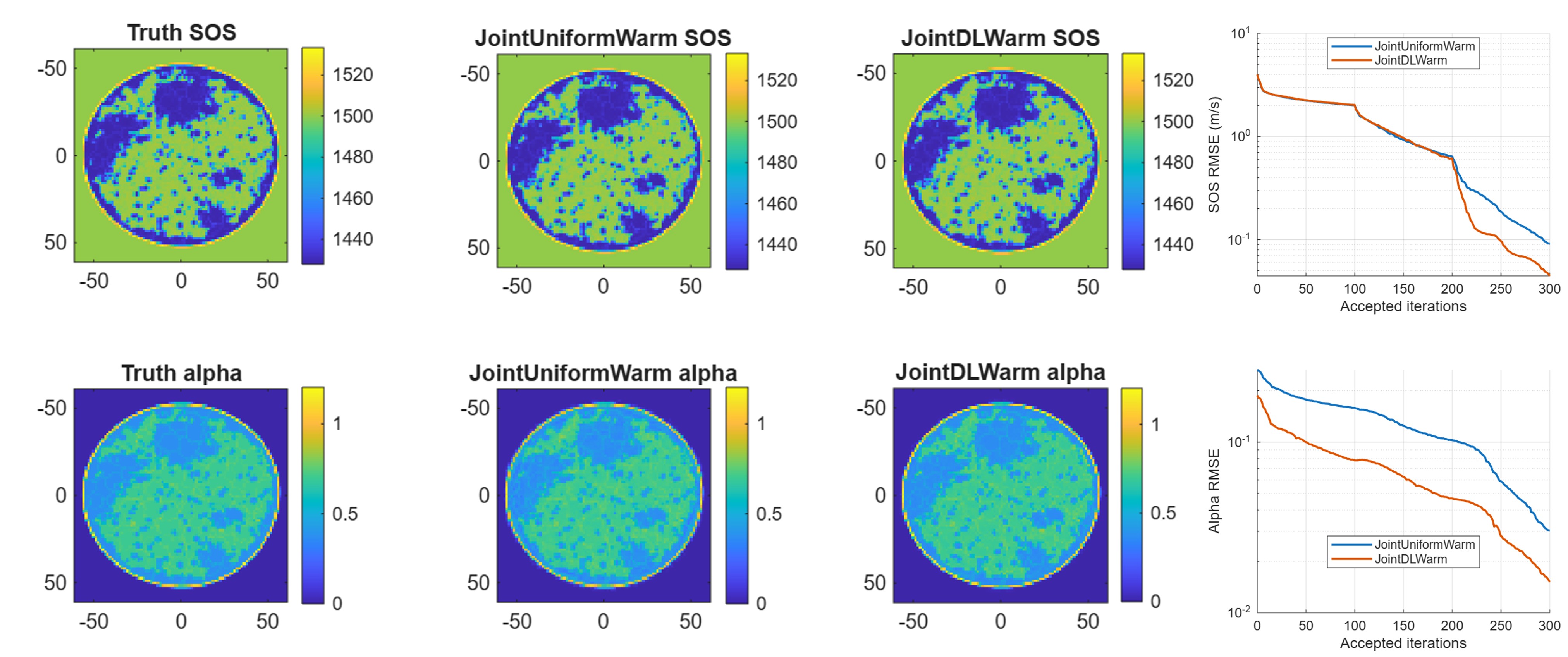}}
\caption{Reconstruction results with different initialization strategies for FWI for experiment B.}
\label{fig:exp3_recon}
\end{figure}

The results from Experiment B are provided in Fig. \ref{fig:exp3_recon}. The images for both attenuation and SOS are of high fidelity, which suggests that the attenuation can be reconstructed accurately if an accurate SOS initialization is used and enough processing time is available. However, the DL attenuation initialization can reduce the computational cost to solve the FWI as shown in Table  \ref{tab:matched_quality_cost}, where values report the cumulative number of forward and adjoint solves at the first recorded iteration, satisfying both the specified attenuation RMSE threshold and a SOS RMSE of $\leq 1$~m/s. The DL initialization reduced the cost by approximately 30\% compared to uniform initialization, and approximately halved both final parameter errors at the same total computational budget.

\begin{table}[!htbp]
\centering
\caption{Computational cost of uniform and DL attenuation
initialization for achieving matched reconstruction quality.}
\label{tab:matched_quality_cost}
\small
\renewcommand{\arraystretch}{1.2}

\begin{tabular*}{\textwidth}
{@{\extracolsep{\fill}}cccc@{}}
\hline\hline
\textbf{Attenuation RMSE threshold} &
\textbf{Uniform} &
\textbf{DL initialization} &
\textbf{Cost reduction} \\
\hline
$\leq 0.15$ & 3528 & 3680 & $-4.3\%$ \\
$\leq 0.10$ & 5152 & 3680 & $28.6\%$ \\
$\leq 0.08$ & 5800 & 3680 & $36.6\%$ \\
$\leq 0.05$ & 6504 & 4544 & $30.1\%$ \\
\hline\hline
\end{tabular*}
\end{table}

\subsection{Experiment C: SOS Preconditioning Study}
\begin{figure}[!ht]
\centerline{\includegraphics[width=\columnwidth]{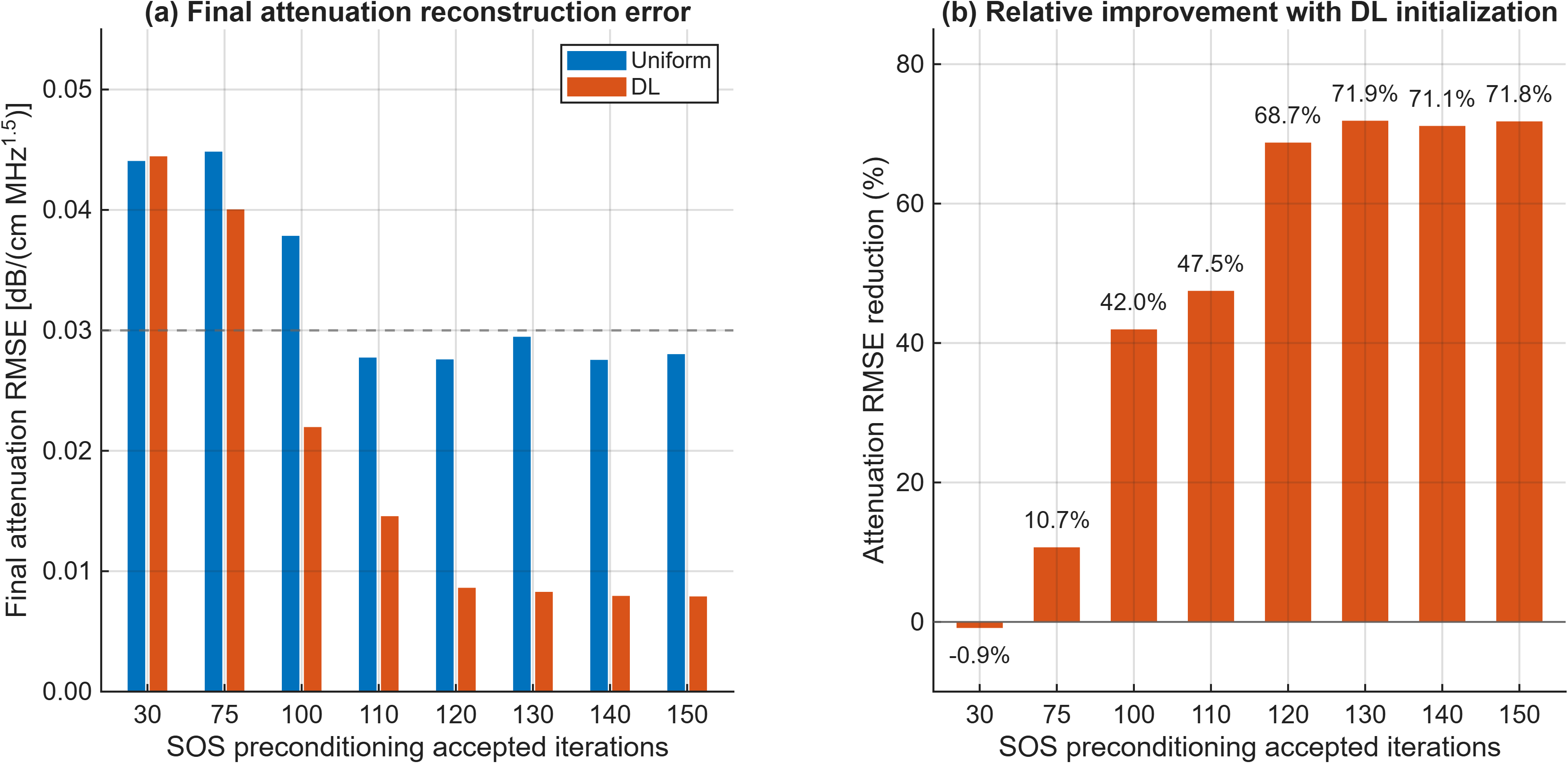}}
\caption{The effect of DL initialization for SOS preconditioning iterations. (a) Final attenuation reconstruction error. (b) Relative improvement with DL initialization.}
\label{fig:exp4}
\end{figure}

The results from Experiment C can be found in Fig. \ref{fig:exp4}. Two things can be concluded from the results. First, DL initialization modestly reduced the minimum tested SOS preconditioning budget from 110 to 100 to meet a prescribed joint reconstruction criterion, i.e., an attenuation RMSE of at most 0.03 dB/(cm·MHz\(^{1.5}\)), a sound-speed RMSE of at most 1 m/s, and a final relative RF residual of at most 0.001.  This means that under the specified joint-inversion budget, DL initialization allowed joint inversion to begin from a less accurate SOS map. Second, the more substantial benefit from DL was improved attenuation accuracy after sufficient SOS preconditioning, rather than a sharp reconstruction threshold. The better the SOS preconditioning is, the higher relative improvement with DL initialization. The relative reduction in attenuation RMSE generally increased with additional preconditioning and reached approximately 69-72\% at the 120-150 iteration checkpoints, as shown in Fig. \ref{fig:exp4} (b).

\section{Discussion}
From these three experiments, it was observed that the DL attenuation initialization could reduce computational cost when the SOS was reconstructed accurately in the first stage. This two stage reconstruction strategy for joint reconstruction of SOS and attenuation in FWI was mentioned in several previous studies \cite{li2026simultaneous,Ali10486959,Liva2017Time}, i.e., first reconstructing SOS and then jointly reconstructing SOS and attenuation. The main reason for this approach is that the SOS is related to phase, while attenuation is related to signal amplitude, which is more sensitive compared to phase change. As a result, an accurate reconstruction of SOS is a prerequisite of jointly reconstructing SOS and attenuation in FWI. In this study, the DL attenuation initialization can save computational cost when the SOS is reconstructed accurately in the first stage. The DL initialization can save computational cost in both the first stage (getting an accurate SOS reconstruction) and the second stage (joint reconstruction of both SOS and attenuation), and will obtain improved reconstruction quality when the first stage SOS reconstruction is better.

\section{Conclusion}
In this study, a deep-learning-assisted FWI for attenuation reconstruction in ultrasound computed tomography was proposed. The result supports the potential of DL initialization to improve the computational efficiency and reconstruction quality of joint SOS and attenuation FWI. DL-based attenuation initialization is an effective strategy for improving attenuation FWI, which saved computational cost by 28\% and improved attenuation reconstruction accuracy by 70\% when combined with sufficiently accurate sound-speed estimates.
%% If only one acknowledgment, 
%% \begin{acknowledgment}...\end{acknowledgment}
%% may be used.
\begin{acknowledgments}
This  work  was  funded  by grants  from  the  National  Institutes  of  Health  (NIH)  (R01CA273700 and R01CA251939)
\end{acknowledgments}
\section*{Author Declarations}
The authors have no conflicts of interest to disclose.
\section*{Data Availability}
The data that support the findings of this study are available from the corresponding author upon reasonable request.
\bibliography{sampbib}

\end{document}